\documentstyle[prd,preprint,aps]{revtex}
\begin{document}
\tightenlines

\draft
\title{Studying the Triple Higgs Self-Coupling Via $e^+e^- \to b\bar b HH, t\bar t HH$
       at Future Linear $e^+e^-$ Colliders}

\author{A. Guti\'errez-Rodr\'{\i}guez $^{1}$, M. A. Hern\'andez-Ru\'{\i}z $^{2}$
        and O. A. Sampayo $^{3}$}

\address{(1) Facultad de F\'{\i}sica, Universidad Aut\'onoma de Zacatecas\\
         Apartado Postal C-580, 98060 Zacatecas, Zacatecas M\'exico.\\
         Cuerpo Acad\'emico de Part\'{\i}culas, Campos y Astrof\'{\i}sica.}

\address{(2) Facultad de Ciencias Qu\'{\i}micas, Universidad Aut\'onoma de Zacatecas\\
         Apartado Postal 585, 98060 Zacatecas, Zacatecas M\'exico.\\
         Cuerpo Acad\'emico de F\'{\i}sico-Matem\'aticas.}

\address{(3) Departamento de F\'{\i}sica, Universidad Nacional del Mar del Plata\\
         Funes 3350, (7600) Mar del Plata, Argentina.}
\date{\today}
\maketitle

\begin{abstract}

We study the triple Higgs self-coupling at future  $e^{+}e^{-}$
colliders energies, with the reactions $e^{+}e^{-}\rightarrow b \bar b HH$ and
$e^{+}e^{-}\rightarrow t \bar t HH$. We evaluate the total cross section of
$b\bar bHH$, $t\bar tHH$ and calculate the total number of events considering
the complete set of Feynman diagrams at tree-level. The sensitivity of the
triple Higgs coupling is considered in the Higgs mass range $110-190$ $GeV$,
for the energy which is expected to be available at a possible Next
Linear $e^{+}e^{-}$ Collider with a center-of-mass energy $800, 1000, 1500$
$GeV$ and luminosity 1000 $fb^{-1}$.

\end{abstract}



\section{Introduction}

In the Standard Model (SM) \cite{Weinberg} of particle physics, there are three
types of interactions of fundamental particles: gauge interactions, Yukawa
interactions and the Higgs boson self-interaction. The Higgs boson \cite{Higgs}
plays an important role in the SM; it is responsible for generating the masses
of all the elementary particles (leptons, quarks, and gauge bosons). However,
the Higgs-boson sector is the least tested in the SM, in particular the
Higgs boson self-interaction.

The search for Higgs bosons is one of the principal missions of present and future
high-energy colliders. The observation of this particle is of major importance
for the present understanding of fundamental particle interactions.
Indeed, in order to accommodate the well established electromagnetic and weak
interaction phenomena, the existence of at least one isodoblete scalar field
to generate fermion and weak gauge bosons masses is required. Despite numerous successes
in explaining the present data, the SM cannot be completely tested before
this particle has been experimentally observed and its fundamental properties
studied.

In the SM, the profile of the Higgs particle is uniquely determined once its
mass $M_H$ is fixed \cite{Hunter}; the decay width and branching, as well as
the production cross sections, are given by the strength of the Yukawa couplings
to fermions and gauge bosons, which is set by the masses of these particles.
However, the Higgs boson mass is a free parameter with two
experimental constraints.

The SM Higgs boson has been searched by LEP in the Higgs-strahlung process,
$e^+e^-\to HZ$, for c.m. energies up to $\sqrt{s}=209$ $GeV$ and with a large
collected luminosity. In the summer of 2002, the final results of the four LEP
collaborations were published and some changes were made with respect to the
original publication. Especially, the inclusion of more statistics, the
revision of backgrounds, and the reassessment of systematic errors. When these results are
combined, an upper limit $M_H \geq 114.4$ $GeV$ is established at the $95 \%$
confidence level \cite{Working Group}. However, in the absence of additional
events with respect to SM predictions, this upper limit was expected
to be $M_H > 115.3$ $GeV$ due to a $1.7\sigma$ excess
[ compared to the value $2.9\sigma$ reported at the end of 2000 ] of events
for a Higgs boson mass in the vicinity of $M_H= 116$ $GeV$ \cite{Working Group}.

The second constraint comes from the accuracy of the electroweak observables
measured at LEP, SLAC Large Detector (SLC), and the Fermilab Tevatron,
which provide sensitivity to $M_H$. The Higgs boson contributes logarithmically,
$\propto\log(\frac{M_H}{M_W})$, to the radiative corrections to the $W/Z$
boson propagators. The status, as found in Summer 2002, is summarized in 
Reference \cite{LEPSLD}. When taking into account all available data,
(i.e. the $Z^0$-boson pole LEP and SLC data, the measurement of the $W$ boson
mass and total width, the top-quark mass and the controversial NuTeV result)
one obtains a Higgs boson mass of $M_H=81^{+42}_{-33}$ $GeV$, which leads to
a 95$\%$ confidence level upper limit of $M_H < 193$ $GeV$ \cite {LEPSLD}.

A detailed study of the Higgs potential represents a conclusive test of
the symmetry breaking and mass generation mechanism. After discovering
of an elementary Higgs boson and testing its couplings to quarks, leptons and
gauge bosons, a further proof of the Higgs mechanism will be the experimental
evidence that the Higgs field potential has the properties required for breaking
the electro-weak symmetry.

The trilinear and quartic Higgs boson couplings \cite{Boudjema,Ilyin,Djouadi}
$\lambda$ and $\tilde\lambda$ are defined through the potential

\begin{equation}
V(\eta_H)=\frac{1}{2}M^2_{H}\eta^2_{H}+\lambda v\eta^3_{H}+\frac{1}{4}\tilde\lambda\eta^4_{H},
\end{equation}

\noindent where $\eta_{H}$ is the physical Higgs field. In the SM, we obtain
$M_H= \sqrt{2\lambda}v$, as the simple relationship between the Higgs boson mass
$M_{H}$ and the self-coupling $\lambda$, where $v=246$ $GeV$ is the vacuum expectation
value of the Higgs boson. The trilinear vertex of the Higgs field $H$ is given
by the coefficient $\lambda_{HHH}=\frac{3M^2_H}{M^2_Z}$ and $M_H$ can be determined
to $O(100 $GeV$)$ accuracy. An accurate test of this relationship may reveal the
extended nature of the Higgs sector. The measurement of the triple Higgs boson
coupling is one of the most important goals of Higgs physics in a future $e^+e^-$
linear collider experiment. This would provide the first direct information
on the Higgs potential that is responsible for electroweak symmetry breaking.

The trilinear Higgs self-coupling can be measured directly in
pair-production of Higgs particles at hadron and high-energy $e^+e^-$ linear
colliders. Several mechanisms that are sensitive to $\lambda_{HHH}$ can be
exploited for this task. Higgs pairs can be produced through double
Higgs-strahlung of $W$ or $Z$ bosons \cite{Gounaris,Ilyin,Djouadi,Kamoshita},
$WW$ or $ZZ$ fusion \cite{Boudjema,Ilyin,Barger,Dobrovolskaya,Dicus}; moreover,
through gluon-gluon fusion in $pp$ collisions \cite{Glover,Plehn,Dawson} and
high-energy $\gamma\gamma$ fusion \cite{Boudjema,Ilyin,Jikia} at photon
colliders. The two main processes at $e^+e^-$ colliders are double
Higgs-strahlung and $WW$ fusion:

\begin{eqnarray}
\mbox{double Higgs-strahlung}&:& e^+e^- \to ZHH  \nonumber \\
\mbox{$WW$ double-Higgs fusion}&:& e^+e^- \to \bar\nu_e \nu_e HH.
\end{eqnarray}

The $ZZ$ fusion process of Higgs pairs is suppressed by an order of magnitude
because the electron-Z coupling is small. However, the process $e^+e^-
\rightarrow ZHH$ has been studied \cite{Gounaris,Ilyin,Djouadi,Kamoshita} extensively.
This three-body process is important because it is sensitive to Yukawa couplings.
The inclusion of four-body processes with heavy fermions $f$, $e^{+}e^{-}\rightarrow f\bar f HH$,
in which the SM Higgs boson is radiated by a $b(\bar b)$ quark at future
$e^{+}e^{-}$ colliders  \cite{NLC,NLC1,JLC,A.Gutierrez} with a c.m. energy
in the range of 800 to 1500 $GeV$, as in the case of DESY TeV Energy Superconducting
Linear Accelerator (TESLA) machine \cite{TESLA}, is necessary in order to know
its impact on three-body mode processes and also
to search for new relations that could have a clear signature of the Higgs
boson production.

The Higgs coupling with top quarks, the largest coupling in the SM,
is directly accessible in the process where the Higgs boson is radiated
off top quarks, $e^{+}e^{-}\rightarrow t\bar t HH$, followed by the process
$e^{+}e^{-}\rightarrow b\bar b HH$. This processes depends on the
Higgs boson triple self-coupling, which could lead us to obtain the first
non-trivial information on the Higgs potential. We are interested in finding
regions that could allow the observation of the $b\bar bHH$ and
$t\bar tHH$ processes at the next generation of high energy $e^{+}e^{-}$ linear colliders.
We consider the complete set of Feynman diagrams at tree-level (Figs. 1, 2)
and use the CALCHEP \cite{Pukhov} packages to evaluate the amplitudes
and cross section.

At the linear collider, the triple Higgs coupling $\lambda_{HHH}$ can be accessed by
studying multiple Higgs production in the reactions
$e^+e^- \to b\bar b HH$ and $t\bar t HH$ that are sensitive to the triple Higgs
vertex. The first process is more relevant at lower values of the center-of-mass
energy $\sqrt{s}$, and for the Higgs boson masses in the range $110 \leq M_H \leq 190$ $GeV$.
The second process is more relevant at collision energies above 1 $TeV$ and
ensures sensitivity to the triple Higgs vertex for intermediate range Higgs masses.

This paper is organized as follows: In Sec. II, we study the triple Higgs
self-coupling through the processes $e^{+}e^{-}\rightarrow b \bar b HH$ and
$e^{+}e^{-}\rightarrow t \bar t HH$ at next generation linear $e^{+}e^{-}$
colliders. In Sec. III, we give our conclusions.

\section{Double Higgs Production Cross Section in the SM at Next Generation
         Linear Positron-Electron Colliders}

In this section we present numerical results for $e^{+}e^{-}\rightarrow b \bar b HH$
and $e^{+}e^{-}\rightarrow t \bar t HH$ with double Higgs production.
We carry out the calculations using the framework of the Standard Model
at next generation linear $e^{+}e^{-}$ colliders. We use CALCHEP \cite{Pukhov}
packages for calculations of the matrix elements and cross-sections. These
packages provide automatic computation of the cross-sections and distributions
in the SM as well as their extensions at tree-level. Both $e^{+}e^{-}\rightarrow b \bar b HH$
and $e^{+}e^{-}\rightarrow t \bar t HH$ processes
are studied, including a complete set of Feynman diagrams. We consider the high
energy stage of a possible Next Linear $e^{+}e^{-}$ Collider with
$\sqrt{s}=800, 1000, 1500$ $GeV$ and design luminosity 1000 $fb^{-1}$.

For the SM parameters, we have adopted the following: the Weinber
angle $\sin^2\theta_W=0.232$, the mass ($m_b=4.5$ $GeV$) of the
bottom quark, the mass ($m_t=175$ $GeV$) of the top quark, and 
the mass ($m_{Z^0}=91.2$ $GeV$) of the $Z^0$, taking the mass
$M_H$ of the Higgs boson as input \cite{Lett}.

\subsection{Triple Higgs Self-Coupling Via $e^{+}e^{-}\rightarrow b \bar b HH$}

To illustrate our results of the sensitivity to the $HHH$ triple
Higgs self-coupling, we show the $\kappa$ dependence of the total
cross-section for $e^{+}e^{-}\rightarrow b \bar b HH$ in Fig. 3. We consider one representative
value of the Higgs mass $M_H=130$ $GeV$ with the center-of-mass energy of $\sqrt{s}= 800,
1000, 1500$ $GeV$ and varying the trilinear coupling  $\kappa\lambda_{HHH}$
within the range $\kappa=-1$ and $+2$. Clearly, the cross-section is sensitive
to the value of the trilinear couplings. Since the $b\bar b HH$ cross-section
and its sensitivity to $\lambda_{HHH}$ decrease with increasing collider energy, a
linear collider operating at 800 $GeV$ offers the opportunity for a precise
measurement of $\lambda_{HHH}$ for $M_H\leq 130$ $GeV$.

Fig. 4 shows the total cross-section as a function of the center-of-mass energy
$\sqrt{s}$ for one value of the Higgs mass $M_H=110$ $GeV$ and for several values
of $\kappa$. We observe in this figure that the total cross-section of $b\bar b HH$
is of the order $0.04$ $fb$ for Higgs mass $M_H=110$ $GeV$ and $\kappa=1.5$.
The cross sections are at the femtobarn fraction level and they quickly
drop with the increase of the center-of-mass energy. Under these conditions,
it would be very difficult to extract any useful information about the Higgs
self-coupling from the studied process unless the $e^+e^-$ machine works
with very high luminosity.

The cross-section for $e^{+}e^{-}\rightarrow b \bar b HH$ as function of the
center-of-mass energy, with $M_H= 130$ $GeV$ and $\kappa=0.5, 1(S.M.), 1.5$ is presented
in Fig. 5. The cross-sections are shown for unpolarized electron and positron
beams. As in Fig. 4, the cross-section is at the femtobarn
fraction level and decreases with rising energy. However, the cross-section
increases with rising self-coupling in the vicinity of the SM value.

We observe in Figs. 4 and 5 that the cross-section decreases as energy increases.
At some given energy, the maximum total cross-section value is $\sigma^{Tot}_{max}$,
depending on the Higgs mass. Since fermion chirality is
conserved at the $Z^0-fermion$ vertex, the cross section may increase by
practically twice when electrons and positrons are polarized. Our conclusion
is that for Higgs masses in the intermediate mass range and rising self-coupling
in the vicinity of the SM value, a visible number of events would be produced
as illustrated in Tables I-III.

For center-of-mass energies of 800-1500 $GeV$ and high luminosity, the
possibility of a detailed study of the triple Higgs boson self-coupling
via the process $b\bar bHH$ is promising as shown in Tables I-III. Thus, a
high-luminosity $e^+e^-$ linear collider is a very high precision machine in
the context of Higgs physics. This precision would allow the determination of
the complete profile of the SM Higgs boson, in particular if its mass is smaller
than $\sim 130$ $GeV$.

\vspace*{5mm}

\begin{center}
\begin{tabular}{|c|c|c|c|}
\hline
Total Production of Higgs Pairs & \multicolumn{3}{c|}{$e^{+}e^{-}\rightarrow b \bar b HH \hspace{8mm} \kappa=0.5$}\\
\hline
\hline
\cline{2-4} & $\sqrt{s}= $ & $\sqrt{s}= $ & $\sqrt{s}= $  \\
$M_H(GeV)$ & 800 $GeV$ & 1000 $GeV$ & 1500 $GeV$ \\
\hline \hline
 110 & 20  & 16  & 10  \\
 130 & 17  & 14  &  9  \\
 150 & 14  & 12  &  9  \\
 170 & 11  & 11  &  8  \\
 190 &  9  & 10  &  8  \\
\hline
\end{tabular}
\end{center}

\begin{center}
Table I. Total production of Higgs pairs in the SM for ${\cal L}=1000$ $fb^{-1}$,
$m_b=4.5$ $GeV$ and $\kappa=0.5$.
\end{center}

\vspace*{5mm}

\begin{center}
\begin{tabular}{|c|c|c|c|}
\hline
Total Production of Higgs Pairs & \multicolumn{3}{c|}{$e^{+}e^{-}\rightarrow b \bar b HH \hspace{8mm} \kappa=1(SM)$}\\
\hline
\hline
\cline{2-4} & $\sqrt{s}= $ & $\sqrt{s}= $ & $\sqrt{s}= $  \\
$M_H(GeV)$ & 800 $GeV$ & 1000 $GeV$ & 1500 $GeV$ \\
\hline \hline
 110 & 23  & 18  & 12  \\
 130 & 21  & 17  & 11  \\
 150 & 18  & 16  & 10  \\
 170 & 15  & 14  & 10  \\
 190 & 13  & 13  & 10  \\
\hline
\end{tabular}
\end{center}

\begin{center}
Table II. Total production of Higgs pairs in the SM for ${\cal L}=1000$ $fb^{-1}$,
$m_b=4.5$ $GeV$ and $\kappa=1(SM)$.
\end{center}

\vspace*{5mm}

\begin{center}
\begin{tabular}{|c|c|c|c|}
\hline
Total Production of Higgs Pairs & \multicolumn{3}{c|}{$e^{+}e^{-}\rightarrow b \bar b HH \hspace{8mm} \kappa=1.5$}\\
\hline
\hline
\cline{2-4} & $\sqrt{s}= $ & $\sqrt{s}= $ & $\sqrt{s}= $  \\
$M_H(GeV)$ & 800 $GeV$ & 1000 $GeV$ & 1500 $GeV$ \\
\hline \hline
 110 & 28  & 21  & 13  \\
 130 & 26  & 21  & 13  \\
 150 & 23  & 20  & 13  \\
 170 & 20  & 18  & 13  \\
 190 & 17  & 17  & 13  \\
\hline
\end{tabular}
\end{center}

\begin{center}
Table III. Total production of Higgs pairs in the SM for ${\cal L}=1000$ $fb^{-1}$,
$m_b=4.5$ $GeV$ and $\kappa=1.5$.
\end{center}

In Fig. 6, we also include a contour plot for the number of events of the
studied process, as a function of $M_H$ and $\sqrt{s}$ with $\kappa=0.5, 1(S.M.), 1.5$.
These contours are obtained from Tables I-III.

\subsection{Triple Higgs Self-Coupling Via $e^{+}e^{-}\rightarrow t \bar t HH$}

As in the case of the process $e^{+}e^{-}\rightarrow b \bar b HH$, in this subsection
we show the sensitivity of the triple Higgs coupling $\kappa \lambda_{HHH}$
via the process $e^{+}e^{-}\rightarrow t \bar t HH$ in the range of $-1\leq \kappa \leq 2$
at tree-level (Fig. 7). Solid (short-dashed, dot-dashed) lines are the results
of the cross-sections of the process $e^{+}e^{-}\rightarrow t \bar t HH$ for
$\sqrt{s}=800, 1000, 1500$ $GeV$ and $M_H=130$ $GeV$. At $\sqrt{s}=1500$ $GeV$,
the cross-section is dominant as illustrated in this figure, where $\kappa=1$
stands for the Standard Model.

Figs. 8 and 9 show the total cross-section as a function of the center-of-mass
energy $\sqrt{s}$ for two representative values of the Higgs mass $M_H=110, 130$ $GeV$,
respectively, and for several values of $\kappa$. We observe in this figure that
the total cross-section of $t\bar t HH$ is of the order of $0.025$ and $0.018$
$fb$ for Higgs masses $M_H=110, 130$ $GeV$, and $\kappa=1.5$.
The cross-sections are at the femtobarn fraction level, and they quickly
drop with the increase of the center-of-mass energy. Under these conditions
(similar to the $e^{+}e^{-}\rightarrow b \bar b HH$ process), it would be very
difficult to extract any useful information about the Higgs self-coupling from
the studied process unless the $e^+e^-$ machine works with very high
luminosity.

We observe in Figs. 8 and 9 that the cross section decreases with energy.
At some given energy, the total cross sections
has its maximum value of $\sigma^{Tot}_{max}$ depending on the Higgs mass. Since
fermion chirality is conserved at the $Z^0-fermion$ vertex, the cross section
may increase by practically twice when electrons and positrons are polarized.

For center-of-mass energies of 800-1500 $GeV$ and high luminosity, the
possibility of observing the process $t\bar tHH$ are promising as shown in
Tables IV-VI. Thus, a high-luminosity $e^+e^-$ linear collider is a very high
precision machine in the context of Higgs physics. This precision would allow
the determination of the trilinear Higgs boson self-coupling of the SM, in
particular if its mass is smaller than $\sim 130$ $GeV$.

\vspace*{5mm}

\begin{center}
\begin{tabular}{|c|c|c|c|}
\hline
Total Production of Higgs Pairs & \multicolumn{3}{c|}{$e^{+}e^{-}\rightarrow t \bar t HH \hspace{8mm} \kappa=0.5$}\\
\hline
\hline
\cline{2-4} & $\sqrt{s}= $ & $\sqrt{s}= $ & $\sqrt{s}= $  \\
$M_H(GeV)$ & 800 $GeV$ & 1000 $GeV$ & 1500 $GeV$ \\
\hline \hline
 110 & 11  & 18  & 17 \\
 130 &  5  & 11  & 13 \\
 150 &  2  &  6  &  9 \\
 170 &  -  &  4  &  7 \\
 190 &  -  &  2  &  5 \\
\hline
\end{tabular}
\end{center}

\begin{center}
Table IV. Total production of Higgs pairs in the SM for ${\cal L}=1000$ $fb^{-1}$,
$m_t=175$ $GeV$ and $\kappa=0.5$.
\end{center}

\vspace*{5mm}

\begin{center}
\begin{tabular}{|c|c|c|c|}
\hline
Total Production of Higgs Pairs & \multicolumn{3}{c|}{$e^{+}e^{-}\rightarrow t \bar t HH \hspace{8mm} \kappa=1(SM)$}\\
\hline
\hline
\cline{2-4} & $\sqrt{s}= $ & $\sqrt{s}= $ & $\sqrt{s}= $  \\
$M_H(GeV)$ & 800 $GeV$ & 1000 $GeV$ & 1500 $GeV$ \\
\hline \hline
 110 & 13  & 21  & 19 \\
 130 &  5  & 13  & 14 \\
 150 &  2  &  8  & 11 \\
 170 &  -  &  -  &  8 \\
 190 &  -  &  -  &  6 \\
\hline
\end{tabular}
\end{center}

\begin{center}
Table V. Total production of Higgs pairs in the SM for ${\cal L}=1000$ $fb^{-1}$,
$m_t=175$ $GeV$ and $\kappa=1 (SM)$.
\end{center}

\vspace*{5mm}

\begin{center}
\begin{tabular}{|c|c|c|c|}
\hline
Total Production of Higgs Pairs & \multicolumn{3}{c|}{$e^{+}e^{-}\rightarrow t \bar t HH \hspace{8mm} \kappa=1.5$}\\
\hline
\hline
\cline{2-4} & $\sqrt{s}= $ & $\sqrt{s}= $ & $\sqrt{s}= $  \\
$M_H(GeV)$ & 800 $GeV$ & 1000 $GeV$ & 1500 $GeV$ \\
\hline \hline
 110 & 15  & 24  & 20 \\
 130 &  6  & 15  & 16 \\
 150 &  3  &  9  & 13 \\
 170 &  -  &  5  & 10 \\
 190 &  -  &  3  &  8 \\
\hline
\end{tabular}
\end{center}

\begin{center}
Table VI. Total production of Higgs pairs in the SM for ${\cal L}=1000$ $fb^{-1}$,
$m_t=175$ $GeV$ and $\kappa=1.5$.
\end{center}

In Fig. 10, we include a contour plot for the number of events of the studied
process as a function of $M_H$ and $\sqrt{s}$ with $\kappa=0.5, 1(S.M.), 1.5$.
These contours are obtained from Tables IV-VI.

Although the Higgs coupling with top quarks, the largest coupling in
the SM, is directly accessible in the process where the Higgs boson is radiated
off top quarks $e^{+}e^{-}\rightarrow t \bar t HH$, the coupling to bottom
quarks is also accessible in the reaction where the Higgs is radiated by a
$b(\bar b)$ quark, $e^{+}e^{-}\rightarrow b \bar b HH$. For $M_H\lesssim 130$ $GeV$,
the Yukawa coupling can be measured with a precision of less than $5 \%$ at
$\sqrt{s}= 800$ $GeV$ with a luminosity of ${\cal L}=1000$ $bf^{-1}$.

Finally, the measurement of the trilinear Higgs self-coupling, which is the
first non-trivial test of the Higgs potential, is possible in the double
Higgs production processes $e^{+}e^{-}\rightarrow b \bar b HH$ and in the
$e^{+}e^{-}\rightarrow t \bar t HH$ process at high energies. Despite their
smallness, the cross sections can be of any practical use for the scientific
community.

\section{Conclusions}

An analysis of triple Higgs couplings will provide a crucial test of the Higgs
mechanism of electro-weak symmetry breaking by directly accessing the shape
of the Higgs field potential. When the study of the 
$e^{+}e^{-}\rightarrow b \bar b HH$ and $e^{+}e^{-}\rightarrow t \bar t HH$
processes can be performed with high accuracy, $e^{+}e^{-}$ linear colliders
represent a possible opportunity for triple Higgs couplings analysis. Examination
of variables sensitive to the triple Higgs vertex and the availability of high
luminosity will allow test the Higgs potential structure
at Future Linear $e^{+}e^{-}$ Collider. Finally, the study of these
processes is important and could be useful to probe the triple Higgs self-coupling
$\lambda_{HHH}$ given the following conditions: very high luminosity, excellent
$b$ tagging performances, center-of-mass large energy and intermediate range
Higgs mass.

\vspace{2.5cm}


\begin{center}
{\bf Acknowledgments}
\end{center}

This work was supported in part by CONACyT and Sistema Nacional de
Investigadores (SNI) (M\'exico). O.A. Sampayo would like to thank CONICET
(Argentina).

\newpage

\begin{center}
{\bf FIGURE CAPTIONS}
\end{center}

\vspace{5mm}

\bigskip

\noindent {\bf Fig. 1} Feynman diagrams at tree-level for $e^{+}e^{-}
          \rightarrow b\bar b HH$.

\bigskip

\noindent {\bf Fig. 2} Feynman diagrams at tree-level for $e^{+}e^{-}
          \rightarrow t\bar t HH$.

\bigskip

\noindent {\bf Fig. 3} Variation of the cross section $\sigma(b\bar b HH)$ with
          the modified trilinear coupling $\kappa\lambda_{HHH}$ at a collider
          energy of $\sqrt{s}= 800, 1000, 1500$ $GeV$ and $M_H= 130$. The
          variation of the cross section for modified trilinear  couplings
          $\kappa \lambda_{HHH}$ is indicated by the solid and dot-dashed
          lines.

\bigskip

\noindent {\bf Fig. 4} The energy dependence of the cross-section with the
          center-of-mass energy $\sqrt{s}$ for a fixed Higgs mass $M_H= 110$ $GeV$.
          The variation of the cross-section for modified trilinear couplings
          $\kappa \lambda_{HHH}$ is indicated by the solid and dot-dashed lines.

\bigskip

\noindent {\bf Fig. 5} The same as in Fig. 4, but for $M_H =130$ $GeV$.

\bigskip

\noindent {\bf Fig. 6} Contours plot for the number of events of the process
          $e^{+}e^{-}\rightarrow b\bar b HH $ as a function of $M_H$ and
          $\sqrt{s}$. The variation of the number of events for modified trilinear
          couplings $\kappa\lambda_{HHH}$ is indicated for $\kappa= 0.5, 1(S.M.), 1.5$.

\bigskip

\noindent {\bf Fig. 7} The same as in Fig. 3, but for the process
          $e^{+}e^{-} \rightarrow t\bar t HH$.

\bigskip

\noindent {\bf Fig. 8} The same as in Fig. 4, but for the process
          $e^{+}e^{-} \rightarrow t\bar t HH$.

\bigskip

\noindent {\bf Fig. 9} The same as in Fig. 5, but for the process
          $e^{+}e^{-} \rightarrow t\bar t HH$.

\bigskip

\noindent {\bf Fig. 10} The same as in Fig. 6, but for the process
          $e^{+}e^{-} \rightarrow t\bar t HH$.

\end{document}